%% file: Revised1.tex
\documentclass[useAMS,usenatbib]{mn2e} 

\usepackage{amssymb}
\usepackage{amsmath}
\usepackage[pdftex]{graphicx,color}
\usepackage{color}
\newcommand{\del}{\partial}
\newcommand{\de}[2]{\frac{\partial#1}{\partial#2}}
\newcommand{\fx}{\frac{\del f}{\del x}}
\newcommand{\ft}{\frac{\del f}{\del t}}
\renewcommand{\fp}{\frac{\del f}{\del p}}
\newcommand{\fy}{\frac{\del f}{\del y}}
\renewcommand{\fm}{\frac{\del f}{\del \mu}}

\newcommand{\dx}{\Delta x}

\newcommand{\dt}{\Delta t}

\newcommand{\Dx}[1]{\frac{\del #1}{\del x}}
\newcommand{\Dt}[1]{\frac{\del #1}{\del t}}
\newcommand{\Dp}[1]{\frac{\del #1}{\del p}}
\newcommand{\Dy}[1]{\frac{\del #1}{\del y}}
\newcommand{\Dm}[1]{\frac{\del #1}{\del \mu}}

\newcommand{\beq}{\begin{align}}
\newcommand{\eeq}{\end{align}}

\newcommand{\Ga}{\Gamma}
\newcommand{\B}{\beta}

\graphicspath{{./}{./paperPlots/}}

\title[A New Time-Dependent Finite Difference Method for Relativistic Shock Acceleration]{A New Time-Dependent Finite Difference Method for Relativistic Shock Acceleration}
\author[S. Delaney et al]{ S. Delaney$^{1,2}$\thanks{E-mail: sdelaney@cp.dias.ie}, P. Dempsey$^{1}$, P. Duffy$^{2}$, T. P. Downes$^{1,3}$\\ $^{1}$School of Cosmic Physics, Dublin Institute for Advanced Studies, 31 Fitzwilliam Place, Dublin 2, Ireland\\ $^{2}$School of Physics, University College Dublin, Belfield, Dublin 4, Ireland\\ $^{3}$School of Mathematical Sciences, Dublin City University, Glasnevin, Dublin 9, Ireland}
\begin{document}

\date{Accepted 2011 November 24.  Received 2011 November 8; in original form 2011 September 22}

\pagerange{\pageref{firstpage}--\pageref{lastpage}} \pubyear{2011} 

\maketitle 

\label{firstpage} 

\begin{abstract} We present a new approach to calculate the particle distribution function about relativistic shocks including synchrotron losses using the method of lines with an explicit finite difference scheme. A steady, continuous, one dimensional plasma flow is considered to model thick (modified) shocks, leading to a calculation in three dimensions plus time, the former three being momentum, pitch angle and position. The method accurately reproduces the expected power law behaviour in momentum at the shock for upstream flow speeds ranging from $0.1c$ to $0.995c$ ($\Gamma \in (1,10]$). It also reproduces approximate analytical results for the synchrotron cutoff shape for a non-relativistic shock, demonstrating that the loss process is accurately represented. The algorithm has been implemented as a hybrid OpenMP--MPI parallel algorithm to make efficient use of SMP cluster architectures and scales well up to many hundreds of CPUs. \end{abstract}


\section{Introduction} 

Particle acceleration at astrophysical shocks has been the subject of a great deal of research for over 30 years. For non-relativistic shocks it has been shown that the diffusion approximation can can be used to find accurate solutions to the governing partial differential equation(s) in a wide range of problems \citep{Dru83}. The diffusion approximation arises from the assumption of near-isotropy of the distribution function in the region around the shock front. However, in the case of relativistic shock fronts this assumption of near-isotropy is no longer valid and it becomes necessary to consider the full angular dependence of the distribution function \citep{Kir99}. This effectively increases the dimensionality of the governing equation. By assuming a gyrotropic distribution this increase is limited to a single dimension in the form of the particle ``direction angle'', i.e. the angle between the particle's mean (gyro-averaged) velocity and the shock normal. For plane parallel shocks the pitch angle is the same as the particle's pitch angle.

Recently, a semi-implicit finite difference, method of lines approach was used to examine time dependent particle acceleration at a thin (discontinuous), non-relativistic shock in the presence of synchrotron and inverse compton losses \citep{Van09}. This work has shown that finite difference methods and the method of lines can be used to great effect to solve the equations of diffusive shock acceleration (DSA) in the diffusion approximation. We would like to dispense with the diffusion approximation and the test particle approximation but, as we will see, this dramatically increases the complexity of the problem and necessitates the use of high performance computing (HPC) techniques.

Perhaps the most successful method that has been used in the analysis of DSA at relativistic shock fronts is the eigenfunction method. First published in the late 1980s \citep{Kir87}, the method was initially used to investigate the spectral index and angular dependence of the distribution function for shocks with upstream Lorentz factors of $\Gamma \lesssim 5$. Only the steady state solution was considered with isotropic, momentum independent diffusion in particle direction angle. The effects of shock structure were also omitted with a discontinuous shock model being employed. By ignoring radiative losses or injection effects and assuming the solution to be a power-law in momentum, the spatial and angular dependence of the solution were then calculated. Soon afterwards \citet{Hea88} investigated momentum dependent, anisotropic diffusion by using the more general Kolmogorov power spectrum for the scattering waves. Following this work, monotonic shock velocity profiles were considered and the assumption of power law solutions was relaxed to allow the investigation of injection effects at low energies \citep{Sch89, Kir89}. The assumption of momentum independent, isotropic diffusion was reintroduced for this work. About 10 years later an additional paper \citep*{Kir00} extended this method to include the full range of shock velocities from non-relativistic to ultra-relativistic. Again, shock structure was neglected and a power law distribution in momentum assumed. Anisotropic diffusion coefficients, independent of momentum were considered, allowing the correlation length of the tangled field parallel to the shock to vary with respect to that perpendicular to the shock. Later high-energy cutoffs due to synchrotron losses were also incorporated by \citet{Dem07}. 

Given the recent success in the application of the method of lines to non-relativistic shock acceleration, it is extended in this work to the relativistic and ultra-relativistic cases. Such a method could might possibly be coupled with a relativistic (magneto)hydrodynamics code to model modified relativistic shocks.
{%\color{blue}
For the present work we partially retain the test-particle approximation, meaning that we will not include the dynamic, time-dependent back-reaction of the accelerated charged particle population on the background plasma. We employ a continuous, steady velocity profile describing the shock which can be chosen as an arbitrary monotonic function of position before the calculation. Therefore a more ``realistic'' modified shock structure could be imposed but we will not propose such a structure here. Moreover, having successfully applied the present methods to a steady shock profile, the introduction of a time dependent shock structure should be relatively straight forward. The additional terms required in the transport equation are, in fact, remarkably similar those containing the spatial dependence of the profile and therefore should not produce significant numerical challenges. Subsequent coupling with a relativistic (magneto)hydrodynamics code could perhaps be considered to study modified shocks. Such a coupling would be far from trivial, involving the inclusion of some complex physical effects and the associated numerical issues. For example, one would have to consider the currents arising from the anisotropic distribution of high energy charged particles and their effect on the plasma flow. The pressure tensor associated with the anisotropic momentum flux could also introduce difficulties. For now we concern ourselves with the feasibility of this numerical approach to the acceleration problem and we leave these extensions for future work.

A number of physical challenges have been presented that call into question the validity of Fermi acceleration at relativistic shocks. In the ultra-relativistic limit we expect shocks to become superluminal with strong cross-field diffusion becoming a necessary ingredient to support the theory (see for example \cite{Kir99}). The detailed structure of the magnetic field and its interaction with the ultrarelativistic plasma flow and high-energy particles about such shocks is far from clear. Promising results have recently been emerging from large particle in cell simulations which can be used to investigate various (micro-)instabilities which can scatter particles and significantly amplify or alter the character of the magnetic field close to the shock \citep{Spi08,Die08,Kes09}. It is clear that the magnetic field about relativistic and ultrarelativistic shocks is quite complex and heavily coupled with the behaviour of the high energy particles produced in such regions. It has been shown that a sufficiently regular magnetic field upstream of an oblique, relativistic shock will cause particles to leave the loss cone due to the Lorentz force and recross the shock having completed $\lesssim 1$ gyration about the upstream field lines \citep{Ach01}. In this case, pitch angle scattering does not describe the particle motion well in the upstream region. Of course the assumption of a sufficiently regular magnetic field to support this process may not be valid due to the various instabilities and turbulence that may exist in the region. In order to proceed with the present method, we will assume that particle transport in the upstream region is well described by the same equation used to model pitch angle diffusion at parallel shocks. We adopt a similar argument to \cite{Kir00}, noting that this equation may well describe particle transport in other more complex cases where particles interact with a more complicated field. In other words, we have assumed that the particles exhibit diffusive behaviour in ``direction angle'' (that between their velocity and the shock normal) rather than pitch angle. Therefore, in this paper, we consider particle acceleration at plane, parallel, relativistic shocks and note that the results may be extensible to other more complex and realistic cases.
}

We will combine a number of the sub-problems tackled separately with the eigenfunction method and seek a numerical solution. Hence we would like to consider the time dependent acceleration of particles at shocks with a wide range of Lorentz factors $(1 < \Gamma \leq 10)$. In addition, we consider the effects of shock structure to permit the investigation of modified shocks, potentially including time-dependent hydrodynamics. We would also like to include anisotropic, momentum dependent diffusion in particle direction angle to permit the introduction of a range of physically motivated diffusion models. Finally, we consider the effects of radiative losses on the spectrum in order to examine the spectral cutoffs produced, which are vital for comparison with observations. It is clear that the complexity of such an algorithm would be significant and would require the use of current HPC methods. We now present an algorithm and associated implementation designed to meet these requirements with the intention of proving the feasibility of this approach and providing a basis for further research.

\section[]{Equation, Discretisation and Numerical Scheme}
\subsection{A relativistic transport equation}\label{eqSec}
We begin with the relativistic transport equation with units chosen such that $c = 1$. Initially we consider isotropic pitch angle diffusion in the co-moving (fluid) frame although this can be relaxed later with only minor adjustments to the model and the code. We will also assume that the hydrodynamical structure of the shock is steady, though a time dependent model could be incorporated. The injection term $Q$ is not explicitly defined here. In this analysis, any change to the injection term will be absorbed into $Q$. The standard relativistic transport equation is expressed in a mixed coordinate system as
\begin{align}\label{relLong}
\Ga (1 + u \mu) \ft & = \Ga (u + \mu) \Dx{u} \Ga^2 \left ( \mu p \fp + (1 - \mu^2) \fm \right ) \nonumber \\
 &+ D(p) \Dm{}\left ( (1 - \mu^2)\fm \right)  - \Ga (u + \mu) \fx \nonumber \\
 &+ g(\mu)\frac{1}{p^2} \Dp{}\left (p^2 L(p) f \right) + Q.
\end{align}
Here the particle momentum $p$ and direction angle $\mu$ are measured in the co-moving (fluid) frame, while time $t$ and position $x$ (and hence the background flow velocity $u(x)$) are measured in the shock rest frame. In general the pitch angle diffusion coefficient $D$ depends on the particle momentum. The energy loss rate $L$ represents radiative losses and we will consider synchrotron radiation in this discussion. Hence $g$ represents the angular behaviour of the synchrotron losses. Transforming the variable $\mu$ to the shock rest frame using
\begin{align} \label{mus}
\mu = \frac{\mu' - u(x)}{1 - \mu' u(x)} \Leftrightarrow \mu' = \frac{\mu + u(x)}{1 + \mu u(x)},
\end{align}
we seek an equation for $\tilde f(t, x, \mu', p)$. Firstly we can write
\begin{align}
\Gamma ( 1 + \mu u) &= \Gamma \left ( \frac{1 - \mu' u}{1 - \mu' u} + \frac{\mu' u - u^2}{1 - \mu' u} \right ) \nonumber \\
 &= \frac{\Gamma}{\Gamma^2 ( 1 - \mu' u)} = \frac{1}{\Gamma(1 - \mu' u)}
\end{align}
and it can be shown that
\begin{align}
(1 - \mu^2) \de{f}{\mu} = (1 - \mu'^2) \de{\tilde{f}}{\mu'}.
\end{align}
From these expressions one can derive the relativistic transport equation with $\mu'$ measured in the shock rest frame, with the appropriate transformation applied to $g$ to give $\tilde g$:
\begin{align}\label{rte}
\Dt{\tilde{f}} + \mu' \Dx{\tilde{f}} &= \Ga^2 \Dx{u} \frac{\mu' (\mu' - u)}{(1 - \mu' u)} p \Dp{\tilde{f}} \nonumber \\
&+ D(p) \Ga^3 (1 - u \mu')^3 \de{}{\mu'} \left ( (1 - \mu'^2) \de{\tilde{f}}{\mu'} \right ) \nonumber \\
&+ g(\mu') \Ga (1 - u \mu') \frac{1}{p^2} \Dp{}\left (p^2 L(p) \tilde{f} \right) + Q
\end{align}
Let $y = \log(p/p_0)$. For synchrotron losses we then have
\begin{align}
L(y) &= L_0 p_0^2 e^{2y} \\
 g(\mu') &= 1 - \mu_{fluid}^2 = 1 - \left ( \frac{\mu' - u}{1 - \mu' u} \right )^2 = \frac{1 - \mu'^2}{\Ga^2 (1 - \mu' u)^2}.
\end{align}
Recovering factors of $c$ where appropriate, and setting $\B = u/c$, we can make the substitution $\tilde{f} = \bar{f} e^{-4y}$ to take advantage of the expected power law spectrum with index $s \sim 4$. Hence $\bar{f}$
\begin{align}\label{rtesync}
\Dt{\bar{f}} + \mu' c \Dx{\bar{f}} &= \Ga^2 \frac{\mu'(\mu' - \B)}{(1 - \mu' \B)} \Dx{\B} c \left ( \Dy{\bar{f}} - 4\bar{f} \right ) \nonumber \\
&+ D(y) \Ga^3 (1 - \B \mu')^3 \Dm{} \left ( (1 - \mu'^2) \de{\bar{f}}{\mu'} \right ) \nonumber \\
&+ \frac{(1 - \mu'^2)}{\Ga(1 - \mu' \B)} L_0 p_0 e^y \Dy{\bar{f}} + Q.
\end{align}

We now consider the simplified case of a momentum independent diffusion coefficient where $D(y) = D_0 \neq 0$ which is also constant across the shock. $D_0$ has units of inverse time, and we can divide equation \ref{rtesync} by $D_0$ to find 
\begin{align}\label{rteOd}
\de{\bar{f}}{\tau} + \mu' \de{\bar{f}}{\xi} &= \Ga^2 \frac{\mu'(\mu' - \B)}{(1 - \mu' \B)} \de{\B}{\xi} \left ( \Dy{\bar{f}} - 4\bar{f} \right ) \nonumber \\
&+ \Ga^3 (1 - \B \mu')^3 \Dm{} \left ( (1 - \mu'^2) \de{\bar{f}}{\mu'} \right ) \nonumber \\
&+ \frac{(1 - \mu'^2)}{\Ga(1 - \mu' \B)} \frac{L_0}{D_0} p_0 e^y \Dy{\bar{f}} + Q
\end{align}
where
\begin{align}
\tau &= D_0 t \\
 \xi &= \frac{D_0 x}{c}.
\end{align} 

We are free to choose any momentum as the reference momentum $p_0$. We could choose the non-relativistic cutoff momentum associated with the synchrotron losses given by
\begin{align}
p_0 = p^* = \left ( \frac{3 L_0 \kappa}{\Delta u} \left ( \frac{1}{u_1} + \frac{1}{u_2} \right ) \right )^{-1}
\end{align}
where $\Delta u = u_1 - u_2$ and the subscripts $1$ and $2$ represent quantities measured upstream and downstream respectively. An expression describing the relationship between the pitch angle diffusion coefficient $D$ and the spatial diffusion coefficient $\kappa$, which is the basis of the diffusion approximation, can be found in \cite{Kir99}. Hence in the non-relativistic limit with $D = D_0 (1 - \mu^2)$ we expect $\kappa = c^2 / (6 D_0)$ so that
\begin{align}
p_0 = \left ( \frac{L_0}{2 D_0 \Delta \B} \left ( \frac{1}{\B_1} + \frac{1}{\B_2} \right ) \right )^{-1}.
\end{align}
where $\B = u/c$. Therefore if we set the unitless parameter
\begin{align}
y_0 = -\log \left ( \frac{L_0 p_0}{D_0} \right ) = \log \left ( \frac{1}{2 \Delta \B} \left ( \frac{1}{\B_1} + \frac{1}{\B_2} \right ) \right )
\end{align}
then we have the non-dimensionalised equation
\begin{align}\label{rtenondim}
\de{\bar{f}}{\tau} + \mu' \de{\bar{f}}{\xi} &= \Ga^2 \frac{\mu'(\mu' - \B)}{(1 - \mu' \B)} \de{\B}{\xi} \left ( \Dy{\bar{f}} - 4\bar{f} \right ) \nonumber \\
&+ \Ga^3 (1 - \B \mu')^3 \Dm{} \left ( (1 - \mu'^2) \de{\bar{f}}{\mu} \right ) \nonumber \\
&+ \frac{(1 - \mu'^2)}{\Ga(1 - \mu' \B)} e^{(y - y_0)} \Dy{\bar{f}} + Q.
\end{align}
Given an expression for the normalised flow speed $\B(x)$ and the injection term $Q$ the solution of this equation can be scaled to find a specific solution for known values of $L_0$ and $D_0$. Furthermore, for $\B(x) \ll 1$, we expect the (non-dimensional) momentum cutoff to occur at $y = 0$.

A very similar transformation can be applied for momentum dependent diffusion. This will be the subject of future work.

\subsection{Notation}
For ease of use we will revert to the symbols $x$ and $t$ for the non-dimensional space and time variables and the transformations outlined above are to be understood in their definition. We will also dispense with the primes and bars on the quantities $\mu'$ and $\bar{f}$. Therefore it should henceforth be understood that \textit{all quantities except for $p$ are measured in the shock rest frame, $p$ being measured in the comoving fluid frame}. We also note that $f$ now describes deviations of the distribution function from a power law of index $s$.

\subsection{Discretisation} The discretisation of the domain in $\mu$ and $y$ that we have chosen is very straightforward. In $y$ a uniformly spaced grid of $O(1000)$ points is used. In $\mu$ we also choose a uniform grid with the spacing chosen to result in a few tens of grid points. The only subtlety involved is that we choose to place the first and last grid points a distance of $\Delta \mu / 2$ inside the geometric end points $\mu = \pm 1$. This choice greatly simplifies the boundary conditions for the chosen numerical scheme, as described in section \ref{bcs}. 

Regarding the $x$ and $t$ domains, two factors must now be accounted for:
\begin{enumerate}
\item The spatial domain must be greater than the diffusion length scale at all (most notably high) momenta.
\item The temporal domain must be greater than the dynamical and acceleration time-scales.
\end{enumerate}
In addition, if we wish to compare our results to those obtained for discontinuous shocks, we should choose the shock width (and hence the spatial grid spacing) to be less than the diffusion length-scale at all (most notably low) momenta. Of course the method is not limited to such cases and in fact this constraint increases the computational complexity of the problem significantly. Let $\ell(y)$ denote the diffusion length-scale at a given momentum and $w$ the shock width. For an explicit finite difference scheme we then have
\begin{align}\label{dynRange}
c \Delta t \sim \Delta x \ll w \ll \ell(y_{min}) \leq \ell(y_{max}) \ll c t_f.
\end{align}
% This set of inequalities highlights the potential difficulties with this approach in terms of dynamic range and computational complexity. The number of iterations required to reach steady state at $t_f$ could be prohibitively large. This problem is worsened when the momentum dependence of the diffusion coefficient leads to $\ell(y_{min}) \ll \ell(y_{max})$. 
%
% As we can see from the inequalities in (\ref{dynRange}), the spatial grid spacing must be sufficiently fine to resolve the shock velocity profile, yet the spatial domain must be large enough to span many diffusion lengths.
These requirements can more easily be met through the use of a non-uniform grid spacing in $x$.
%The smallest spacing, near the shock, is then chosen based on the shock width $w$. At a large distance from the shock, for example $10 \ell(y)$, we only require that $\Delta x \ll \ell(y)$. Therefore, outside of the immediate vicinity of the shock, we can set $\Delta x \sim x / A$ where $A \gg 10$. This will guarantee that the grid spacing is less than one diffusion length at all points in the domain of interest (ie. those that have a significant influence on the shock: $|x| < 10 \ell(y)$). In the immediate vicinity of the shock $(|x| \lesssim w)$, a uniform grid spacing of $w/C, C \gg 1$ is used.

One of the first tests of any code investigating DSA is the production of a power law with the appropriate spectral index. For a strong, non-relativistic shock (compression ratio of $4$), with a momentum independent diffusion coefficient we expect a spectral index of $-4$. It has been shown \citep{Kru94} that a finite difference scheme modeling a continuous velocity profile will only produce the correct power law if
\begin{enumerate}
\item the advection length-scale per time step is less than the shock width and
\item the diffusion length-scale per time step is greater than the shock width.
\end{enumerate}
The first of these conditions is trivially met for explicit finite difference schemes. The second condition requires consideration when choosing the ratio $C$ of the shock width to the grid spacing at the shock. For a fixed shock width we can proceed to increase $C$ by trial and error until the spectral index converges. A useful ``rule of thumb'' can be found by combining the above inequalities resulting in the requirement that
\begin{align}
w_0 < \frac{2 \dt \beta}{C \dx}
\end{align}
where $w_0$ is the shock width in units of the diffusion length-scale. Hence as a starting point we can choose the number of points resolving the shock and then adjust the shock width to meet this condition. The situation is less clear for relativistic flow where the diffusion length-scale is not well defined.

\subsection{Numerical scheme}

Finite difference methods are one of the most commonly used numerical approaches to solving partial differential equations. After years of development, a wide variety of schemes now exist to match the wide spectrum of problems they are used to tackle. In general, the application of finite difference methods to a given equation requires careful selection of an appropriate approximation scheme from a relatively short list of commonly used ones. It is often instructive to start with the simplest, most crude schemes before deciding which of the more complex and accurate schemes could yield the most efficient results. We use the method of lines here, meaning that we will replace all derivatives \textit{w.r.t.} $\{x,y,\mu\}$ with finite difference approximations and solve the resulting equation for $\ft$ to advance the solution in time. This solution can be approximated by taking discrete steps forward in time using a standard O.D.E. method. We have implemented the ubiquitous fourth order Runge--Kutta method, known for its reliability, and a third order Adams--Bashforth method.

The simplest approach to the diffusion term is the standard centred difference approximation to the second derivative which is second order accurate in $\Delta \mu$. The source terms, containing no derivatives, do not require a difference approximation and are trivially implemented in the scheme. The only remaining terms are advective, representing the uniform motion, shock acceleration and radiative losses (deceleration) of particles.

{%\color{blue}
Because the MPI parallelisation has been implemented in the $y$ dimension, we use the Lax-Wendroff scheme in $y$ to minimize communication costs while giving second order accuracy.
%Unfortunately this does produce some unwanted oscillatory behaviour near the boundaries and for some scenarios the Beam--Warming scheme may be preferable for that reason. In the $x$ direction the additional accuracy of the Lax-Wendroff scheme is desirable, but the CFL condition is more restrictive. Fromm's method is effectively an average of the Beam--Warming and Lax-Wendroff schemes. It produces more accurate results and permits the same stable time step as the Beam--Warming scheme but it uses a four point stencil. However, for the same computational cost, w
A third order scheme is used in $x$ to provide appropriate accuracy for the steep gradients near the shock without incurring excessive computational cost. It behaves well on the non-uniform grid and while high order schemes can be unstable (especially where large gradients are present), this one seems well suited to our particular problem. The scheme for a non-uniform grid is given in the appendix.}

In summary, the numerical approximations used for each derivative in equation \ref{rtenondim} are therefore the following \begin{description} \item $\ft$ : The third order Adams--Bashforth method (explicit, multi-step), or fourth order Runge--Kutta (multi-stage). \item $\fx$ : The third order, upwind-biased scheme (see appendix). \item $\fy$ : The Lax-Wendroff scheme (second order). \item $\Dm{} \left ( (1 - \mu^2) \fm \right )$ : Central differencing (second order). \end{description} 

\subsection{Boundary conditions}\label{bcs} The final requirement for the implementation of this scheme is a set of boundary conditions. Far downstream of the shock we expect the spatial gradient of the distribution function to be close to zero. Furthermore, any errors incurred there should have a minimal effect on the solution near the shock. The upstream boundary is more critical at the shock due to the bulk flow direction. The distribution function decays exponentially as we move upstream from the shock and, far upstream, tends to zero. Thus the boundary condition should approximate either an exponential decay or could be set to zero if sufficiently distant from the shock. 

Suitable boundary conditions in pitch angle (cosine) are perhaps less obvious. In early testing a zero gradient condition was used, accurate to first order, at $\mu = \pm 1$. This did produce reasonable results, comparing well to the expected distributions except at the boundary points. While a second order accurate, zero gradient condition produced slightly better results a more elegant solution was subsequently found. By choosing the grid points to lie at $\{ -1+\Delta \mu/2, -1+3 \Delta \mu / 2, \ldots, 1 - \Delta \mu/2 \}$ in conjunction with the standard central difference approximation to $\Dm{} \left ( (1 - \mu^2) \fm \right )$ the values of the distribution function at $\mu = \pm 1$ conveniently vanish from the calculation. Thus no boundary condition is in fact required. Of course a boundary condition can be chosen, \textit{a posteriori}, and fitted to the data if more detail near the end points is ever required. Alternatively we could extrapolate a suitable fitting function (see section \ref{results} for an example) to evaluate the solution very close to $\mu = \pm 1$. The extrapolation distance can be reduced with higher resolution given sufficient computational resources. 

Finally, we must choose boundary conditions in momentum (magnitude). In the presence of losses it is clear that the solution will decay to zero in the limit of large momentum. Since the shape of the solution in the cutoff region is not generally known it is difficult to make any further assertions without reducing their applicability to special cases. Thus the upper boundary condition is chosen to lie well above the cutoff where a function value of zero is enforced. Excluding points close to the shock, the lower boundary condition in momentum is somewhat less important because loss processes dominate any minimal acceleration and cause particles to flow outwards across the boundary. Indeed, testing has shown the lower boundary condition in momentum away from the shock to have negligible effect. The situation is somewhat complicated close to the shock where (for most pitch angles) acceleration can overcome losses leading to an inflow of particles across the boundary. This inflow of particles is effectively an injection process, representing the point at which particles enter the domain. Since the detail of the injection process is not generally known, we have chosen to implement a constant source of particles at the shock with a sink (zero boundary condition) elsewhere on the lower momentum boundary (which are of little significance to the solution). 

\section[]{Domain of interest and computational parallelism} The first task in the implementation of a finite difference scheme is to establish the domain and resolution of the numerical grid. In the $x$ direction, the domain must enclose the region of plasma that can significantly affect the distribution at the shock. Hence the downstream boundary is chosen to lie at $G \ell_2(y)$ where $G \gg 1$ and the subscript $2$ denotes a downstream quantity. The diffusion length-scale in the non-relativistic case is given by $\ell_2(y) \simeq \kappa(y) / u_2$ where $\kappa$ is the spatial diffusion coefficient and $u_2$ the downstream flow speed. Beyond the boundary position, particles are sufficiently unlikely to diffuse back to the shock that they can be neglected. In the upstream region the solution function is expected to decay exponentially with increasing distance from the shock. We therefore set our upstream boundary at a distance such that the particle phase space density has decayed to a negligible amount. A numerical value for the approximate exponent associated with this decay can be found from the useful expressions given in \citet{Dem08}.

The $y$ domain can be chosen based on the estimated value of the momentum cutoff $p_0$, which implies a cutoff near $y = 0$. Based on previous work such as \cite{Dem07}, we can anticipate that the cutoff should take place over $2$ or perhaps $3$ decades in momentum. Hence, in order to examine the cutoff and power law region, we could choose the domain $y \in [-10, 3]$. Of course this estimate can be revised if it becomes apparent that the necessary features are not captured. If we inject particles in a narrow energy range, it is also necessary to leave some room at the low energy end of the spectrum for the oscillations described in \cite{Kir89} to decay. 

The domain in $\mu$ must obviously span the range $(-1, 1)$. The resolution must be sufficient to resolve any features in the distribution, but this scale is not immediately obvious a priori. We can, of course, run the code and decrease $\Delta \mu$ as necessary for adequate resolution. Even in the case of high velocity flow, it appears that only a few tens of points are necessary for sufficient angular resolution. Of course this may not be the case if $\mu$ is measured in a different frame. 

In order to attain more reasonable run times, parallelisation using MPI is necessary. The grid has largest dimension in the $y$ direction and hence we implement MPI parallelisation in that dimension. We use the Stokes cluster, run by the Irish Centre for High End Computing (ICHEC), which consists of $320$ hex-core SMP nodes with ConnectX infiniband interconnects. Non-blocking communications were employed, as well as parallel data output, using the \textit{MVAPICH} MPI implementation.

Improved scalability can be attained on such SMP clusters through the use of a hybrid OpenMP--MPI implementation. This is achieved by reducing the number of MPI processes in operation (which reduces the relative size of the communication halo and hence the communication overhead) and running numerous OpenMP threads per MPI process on each SMP node. Thus running $N$ threads per process we reduce the relative size of the halo by a factor of $N$. The total factor by which parallelisation overhead is reduced depends on the OpenMP overhead incurred, but for many problems it can be reduced by a factor of order $N$.
%For this particular application, the loop ordering was chosen such that the outermost loop was that which typically had the largest number of elements. This permitted efficient parallel execution of the outer loop by OpenMP with the finest possible load balancing. Note that the instructions executed in each OpenMP thread then has a smaller number of elements in the outer loop by a factor of $N$. Also note that using OpenMP parallelisation on the outer loop only incurs the initialisation overhead (ie. creating a team of threads, scheduling, etc) once per time step. The buffering associated with the communication and the enforcement of boundary conditions were also executed in parallel using OpenMP with the ``longest loop outside'' strategy.

%The remaining (inner) loops were ordered in the standard fashion, long loops inside short ones, for highest efficiency in serial execution. The memory allocation was implemented such that sequential elements of the innermost loop were contiguous in memory, minimising pointer overhead and simplifying the caching of sequentially accessed data. It is also possible to utilise Intel's hardware multithreading (``Hyper-threading'') by executing 12 threads on each hex-core node. In this configuration pairs of threads share the same memory bandwidth and hence it proves optimal to run 2 MPI processes, each running 6 threads, on every hex-core node. 

\section[]{Results}\label{results} 

\begin{figure}
 \centering
 \resizebox{0.45\textwidth}{!}
 { 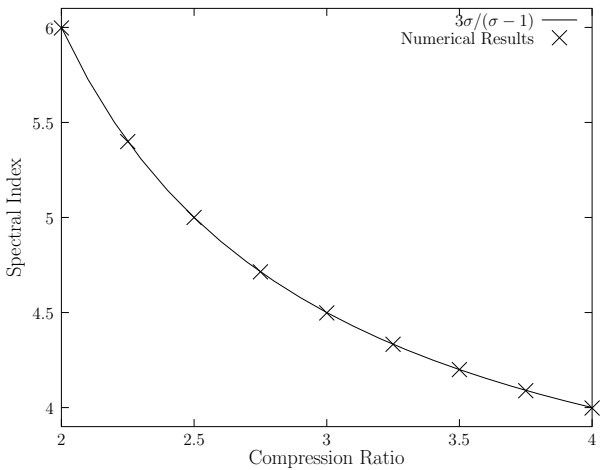 }
\small \caption{Spectral Index vs Compression Ratio for a shock with upstream flow speed $0.1c$. The line represents the non-relativistic analytical result in the diffusion approximation. The crosses mark the numerical results tested.} \label{specNR}
\end{figure}

The primary test used in the development of this method was the production of a power law with appropriate spectral index. If we temporarily ignore the underlying physics that produces a particular compression ratio, we can investigate the spectral indices produced at a non-relativistic shock. We should find that the spectral index is given by the expression $s = 3 \sigma / (\sigma - 1)$ where $\sigma$ is the compression ratio. For testing purposes we chose an upstream flow speed of $u_1 = 0.1c$ with an isotropic, momentum independent diffusion coefficient. No radiative losses were included for this test. The results are shown in Figure \ref{specNR} where we can see the method accurately reproduces the expected results.

%\begin{center} %\begin{tabular}{lll} %hline %$\sigma$ & $s$ & $\Delta s / s$ \\ %\hline %$4$ & $3.998$ & $5\E{-4}$ \\ %$3.75$ & $4.089$ & $5.3\E{-4}$ \\ %$3.5$ & $4.200$ & $2.7\E{-5}$ \\ %$3.25$ & $4.333$ & $2.3\E{-5}$ \\ %$3$ & $4.498$ & $4.4\E{-4}$\\ %$2.75$ & $4.714$ & $4.2\E{-5}$\\ %$2.5$ & $5.000$ & $8.0\E{-5}$\\ %$2.25$ & $5.399$ & $1.5\E{-4}$\\ %$2$ & $5.997$ & $5.0\E{-4}$ %\end{tabular} %\end{center} 

\begin{figure}
 \centering
 \resizebox{0.45\textwidth}{!}
 { 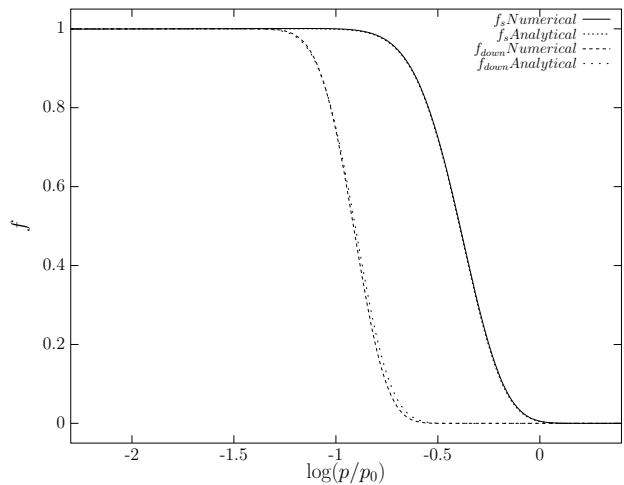 }
 \small \caption{The steady state distribution in momentum near a non-relativistic shock of compression ratio $4$ integrated over pitch angle (a) at the shock and (b) $\sim 10$ diffusion lengths downstream. Momentum independent, isotropic pitch angle diffusion and synchrotron losses used for comparison with \citet{Hea87}.} \label{H+M}
\end{figure}

Figure \ref{H+M} compares the results of our method to analytical work by \cite{Hea87} in the case of a non-relativistic shock with momentum independent diffusion. While the analytical results were obtained using the diffusion approximation, we can see that the results are quite accurate and it is difficult to differentiate between the curves in each pair above the injection energy. In this case the compression ratio of the shock was set at $4$ resulting in a spectral index of $4.001$ between the injection and cutoff energies. Hence the relative error in the spectral index produced by this method is less than $0.1 \%$ for the grid resolution used. The synchrotron cutoff region shows the strong agreement between our numerical results and the analytical ones of \citet{Hea87}. There is a slight discrepancy between the numerical and analytical results downstream, the numerical cutoff being slightly sharper. This may be the result of the increasing spatial grid spacing. The results shown use a magnetic field proportional to the flow velocity and a diffusion coefficient that is constant in space.

Near the injection energy the method also produces the oscillatory behaviour elegantly explained in \cite{Kir89}, corresponding to the various generations of particles recrossing the shock after initial monoenergetic injection. Because the energy gain is pitch angle dependent for each interaction with the shock, the peaks in the spectrum spread out and settle down to a smooth power law. This injection region is not our primary interest in this study and hence we have not included the necessary mathematical details to represent particles with speeds significantly less than $c$. Therefore we will not analyse this oscillatory behaviour any further, other than to note that some dampened oscillations are present at low energies as expected. %, which corresponds to a separation of $\sim 0.95$ in Figure \ref{H+M}.

\begin{figure}
 \centering
 \resizebox{0.45\textwidth}{!}
 { 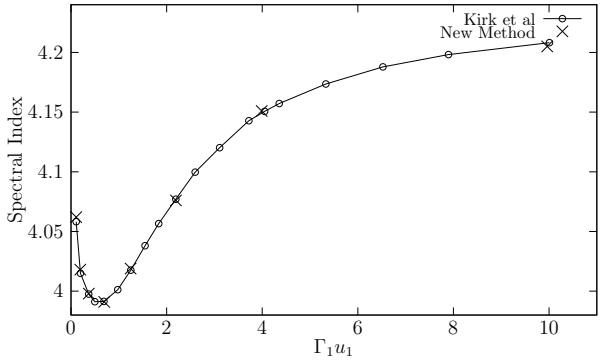 }
 \small \caption{Spectral index at a weakly magnetised thin shock compared with results from \protect\cite{Kir00} for a range of shock speeds.} \label{GUvS}
\end{figure}

For relativistic shocks it is difficult to find purely analytical results for testing purposes. We turn to the eigenfunction method for the most accurate results available that can provide useful tests for our code. Once again, our first test is a check on the spectral indices produced where radiative losses are unimportant. We have employed the J\"{u}ttner-Synge equation of state in order to calculate the compression ratio. A similar scenario was considered by \citet{Kir00}, where the effects of magnetisation were also taken into account in the compression ratio. Using \textsc{Dexter}, a data extraction applet provided by the Astrophysics Data System (ADS), reasonably accurate numerical values can be obtained for comparison with our results. Figure \ref{GUvS} compares the values extracted with those calculated using our method. It is clear that the values obtained for the spectral index are consistent, following this relatively complex curve quite effectively. Of course a very precise comparison would require more accurate knowledge of the results from the eigenfunction method and significant computational resources for higher resolution in our numerical results.

Regarding strongly relativistic shocks, we note that as the Lorentz factor of shock increases the rates of diffusion and acceleration increase rapidly. This introduces stiffness in the equation, the timestep being severely restricted by the factor of $\Gamma^3$ in the diffusion term. Ideally a new numerical scheme should be constructed to deal with this problem. A split method using an implicit scheme or ``super-timestepping'' for diffusion could potentially be used. Depending on the thickness of the shock profile, the $\Gamma^2$ factor in the acceleration term might also lead to stiffness, requiring additional splitting. The present implementation does not employ such methods, but works well for Lorentz factors up to $10$ or more.

\begin{figure}
 \centering
 \resizebox{0.45\textwidth}{!}
 { 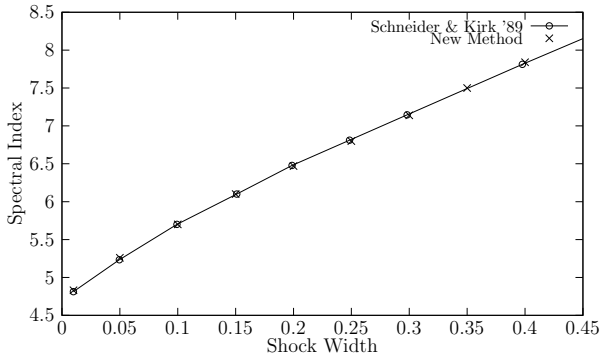 }
 \small \caption{Spectral index at a shock of speed $0.9c$ and compression ratio $2.43$ compared with results from \protect\cite{Sch89} for a range of shock widths. Non-dimensional units are shown, as outlined in section \protect\ref{eqSec}.} \label{WvS}
\end{figure}

We would also like to verify that the method produces the expected behaviour at thick shocks. In order to do this we can compare the spectral index produced using our method with those produced by the eigenfunction method for thick shocks. \cite{Sch89} investigated this effect by assuming a power law solution without radiative losses in the steady state. Once again, we have used the \textsc{Dexter} data extraction applet to compare results from that paper with our numerical work in Figure \ref{WvS}. A shock of speed $0.9c$ and compression ratio $2.43$ is considered for a range of shock widths $w$. While the precision of the comparison is limited by the accuracy of the data extraction, we can see that the results agree quite well even for these expedient, low resolution tests. We should mention that our implementation uses a error function expression to represent a sigmoid velocity transition at the shock instead of the hyperbolic tangent expression used by \cite{Sch89}. Qualitatively the two are very similar. Only a slight adjustment in the units was required to generate the plot shown in Figure \ref{WvS}, and so we did not bother to change our shock profile for this comparison.

Another test can be performed examining the angular distribution of particles at the shock. It has been shown that the first eigenfunction in the expansions used by \cite{Kir00} provides a reasonable approximation at relativistic shocks. Indeed, a good deal of further analysis can be performed using a single eigenfunction approximation as shown in \citet{Dem08}. With pitch angle measured in the shock frame and momentum (magnitude) measured in the fluid frame \cite{Kir00} give \begin{align} f_s \propto \exp \left (- \frac{1 + \mu}{1 - \mu \beta_1} \right ) \end{align} while \citet{Dem08} give \begin{align} f_s \propto \exp \left (a_u \frac{\mu - \beta_1}{1 - \mu \beta_1} \right ) \propto \exp \left (a_u (1 - \beta_1) \frac{1 + \mu}{1 - \mu \beta_1} \right ) \end{align} where $a_u$ is the solution of the transcendental equation $\exp(2a_u) = (a_u(\beta-1)-1)/(a_u(\beta+1)-1)$ (easily approximated numerically). Clearly the two eigenfunctions will be similar if $a_u \simeq (\beta_1 - 1)^{-1}$. Using a numerical root finding algorithm it is straightforward to compare the two and show that the fractional difference falls below $10\%$ near $\beta_1 = 0.5c$, decreasing rapidly for faster shocks. Indeed, \citet{Kir00} state that their first eigenfunction (in isolation) only provides a good approximation above $0.5c$ whereas the eigenfunction used by \citet{Dem08} yields accurate results at lower velocities. 

\begin{figure}
 \centering
 \resizebox{0.45\textwidth}{!}
 { 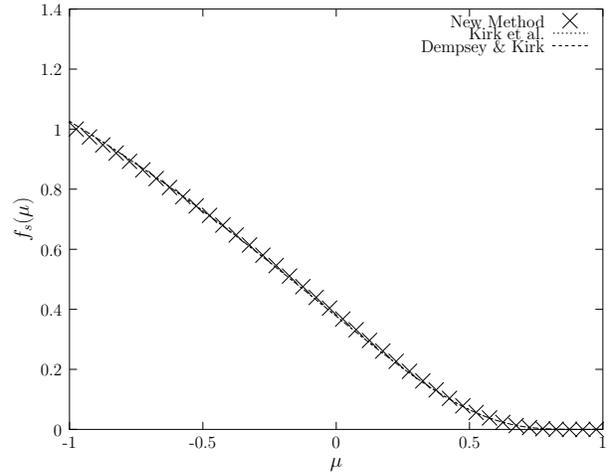 }
 \small \caption{Angular distribution at shock for $u_1 = 0.9c$} \label{FvMu}
\end{figure}

\begin{figure}
 \centering
 \resizebox{0.45\textwidth}{!}
 { 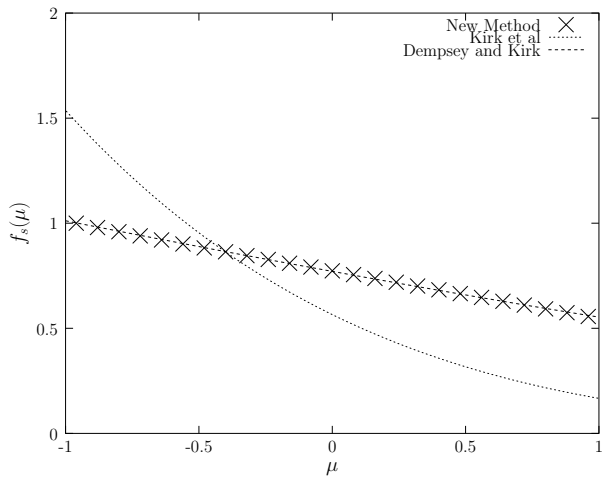 }
 \small \caption{Angular distribution at shock for $u_1 = 0.1c$} \label{FvMu2}
\end{figure}

Figure \ref{FvMu} shows the angular dependence of the distribution function at a shock of upstream flow speed $0.9c$. It is clear that the two eigenfunction solutions are nearly identical in this case. We can see that our numerical method produces a very similar angular spectrum even from a low-resolution grid ($\Delta \mu \simeq 0.143$ in this case). Figure \ref{FvMu2} shows the angular distribution at a shock of upstream flow speed $0.1 c$. Here we can clearly see that the eigenfunction of \citet{Dem08} retains its accuracy at low velocities and agrees well with the numerical results. We expect the solution to become isotropic in the fluid rest frame at large distances downstream of the shock and this is also confirmed in the numerical results. Note that the end points of the numerical domain in $\mu$ lie within half of the grid spacing of the physical extrema $\mu = \pm 1$. Hence, for our numerical scheme, the vanishing factor of $1 - \mu^2$ makes it unnecessary to evaluate any derivatives at $\mu = \pm 1$, thus removing the need for boundary conditions. Note that we are not imposing the condition that there are no particles at $\mu = \pm 1$. If a boundary condition were necessary for a different scheme, a reflecting condition would be appropriate.


\begin{figure}
 \centering
 \resizebox{0.45\textwidth}{!}
 { 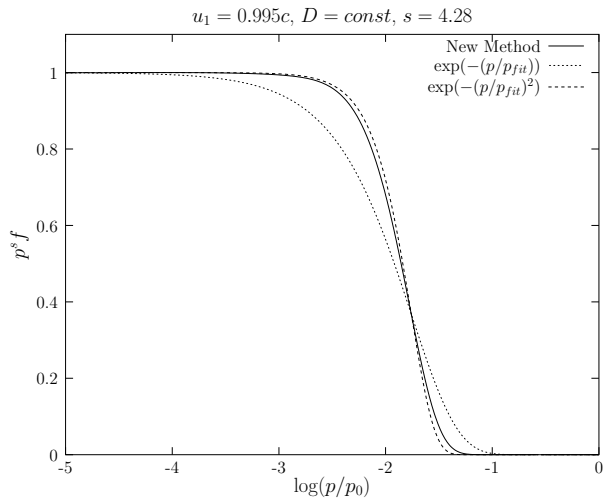 }
 \small \caption{Particle spectrum for an ultrarelativistic shock of width $10^{-3}$ (see \ref{eqSec} for units).} \label{relCut}
\end{figure}

Finally we show an example of a particle spectrum produced by our method to demonstrate its use. Figure \ref{relCut} shows the spectrum of high energy particles produced at a shock of speed $0.995 c$ ($\Gamma \simeq 10$) in the presence of synchrotron losses. The synchrotron loss rate used in the upstream region differs from that used downstream by the square of the shock compression factor, which is approximately $3$. The diffusion coefficient is constant in both space and momentum, rescaled to $1$. The shock velocity transition is based on the error function on a length-scale of $10^{-3}$ in the units described in section \ref{eqSec}, that being about $8$ times longer than the decay length-scale of the solution upstream. Hence we are not modelling a ``thin'' shock here. The spectral index $s$ in the power-law region (well below the cutoff) is approximately $4.28$. Also shown for comparison are two curves describing spectra of the form $f \propto p^{-s} exp(-(p/p_{fit})^\beta)$ for $\beta = 1,2$ where $p_{fit}$ was chosen to match the cutoff position.

\section[]{Conclusion}\label{conclusion} We have presented a new approach for the solution of the particle transport equation about thick, hydrodynamically steady, plane relativistic shocks in one spatial dimension. This approach uses the fluid approximation to model evolution of the phase-space density function of the high energy particles using finite difference approximations and the method of lines. Our objective is to combine and extend the results obtained with the eigenfunction method by various authors such as \cite{Kir87}, \cite{Hea88}, \cite*{Kir00}, \cite{Dem07}. Another significant benefit of the finite difference method of lines approach lies in the ease with which it could be coupled with a relativistic hydrodynamics code to model modified relativistic shocks (with time dependent structure). Our physical model extends that of \citet{Kir00} with the addition of time dependence, a sigmoid shock velocity profile (for first order Fermi acceleration) and synchrotron losses. We have chosen a modified mixed coordinate system in which all quantities are measured in the shock rest frame except for $p$, the magnitude of the particle momentum, which is measured in the rest frame of the background (thermal) fluid flow. The angular distributions produced using this coordinate system are quite smooth, which is numerically advantageous.

The numerical scheme we have chosen is a fourth order Runge--Kutta integrator in conjunction with a second order finite difference scheme in momentum and pitch angle and a third order scheme in space. A non-uniform grid is used in space to focus the computational effort about the shock. The scheme is implemented using a parallel, hybrid (MPI + OpenMP) algorithm to take advantage of modern SMP cluster architectures, demonstrating good scalability up to many hundreds of CPUs.

The method has been tested using analytical results in the non-relativistic limit. As expected, power law solutions are produced with spectral indices agreeing very well with standard theory for a range of compression ratios. At the synchrotron cutoff, the particle spectrum confirms the analytical approximations of \citet{Hea87}, both at the shock and in the region around it. The angular distribution is close to isotropy in non-relativistic cases, as we would expect, quantitatively reproducing analytical results. For relativistic shocks this method also produces power law solutions. The associated spectral indices can be compared with the numerical results obtained using the eigenfunction method, shown in \citet{Kir00}. We have demonstrated that our method accurately reproduces these spectral indices from non-relativistic to ultra-relativistic shocks. In all cases the angular distribution of particles was found to be in agreement with analytical approximations based on single eigenfunction solutions \citep{Dem08}.

We have shown that standard CFD techniques can be used to find numerical solutions of the transport equation for high energy particles in the vicinity of relativistic shocks. A number of physical effects have been incorporated and it is possible, if not straight forward, to include additional effects using these highly versatile finite difference methods. In addition to the results that can be obtained directly from this code, the methods employed would lend themselves particularly well to combined hydrodynamical models in order to investigate acceleration at time dependent, modified relativistic shocks.

\section*{Acknowledgements}
S. Delaney would like to sincerely thank:
\begin{itemize}
\item[-] F. Aharonian, L. O'C. Drury and P. Duffy for the supervision of S. Delaney's PhD studies.
\item[-] T. P. Downes and K. Rochford for their invaluable advice on the computational science associated with this work.
\item[-] The Irish Centre for High End Computing and the e-Inis project for granting us the computational resources required to conduct this research and Science Foundation Ireland for funding S. Delaney's PhD. studies.
\item[-] {%\color{blue}
The referee for reviewing this paper so promptly and suggesting some important improvements.}
\end{itemize}

\appendix
\section{Finite Difference Approximation Schemes}
% The Beam--Warming scheme for the constant velocity advection equation on a non-uniform grid can be written
% \begin{align}
% f_i^{n+1} = f_i^n &+ u \tau \left ( \frac{u \tau - h_1}{h_2(h_1+h_2)} f_{i-2}^n \right . \nonumber \\
%  &\left . + \frac{h_1 + h_2 - u \tau}{h_1 h_2} f_{i-1}^n - \frac{2h_1+h_2 - u \tau}{h_1(h_1+h_2)} f_{i}^n \right )
% \end{align}
% where $h_{1,2}$ are the nearest and next nearest grid spacings in the upwind direction.

The third order scheme we have implemented can, for the constant velocity advection equation, be written
\begin{align}
f_i^{n+1} &= f_i^n + u \tau \frac{u^2 \tau^2 + (h_2 - h_1) u \tau - h_1 h_2}{(h_0 h_1 + h_0^2) h_2 + h_0 h_1^2 + 2 h_0^2 h_1 + h_0^3} f_{i - 2}^n \nonumber \\
 &- u \tau \frac{u^2 \tau^2 + (h_2 - h_1 - h_0) u \tau + ( - h_1 - h_0) h_2}{h_0 h_1 h_2 + h_0 h_1^2} f_{i-1}^n \nonumber \\
 &+ u \tau \left ( \frac{u^2 \tau^2 + (h_2 - 2 h_1 - h_0) u \tau}{(h_1^2 + h_0 h_1) h_2} \right . \nonumber \\
 &+ \left . \frac{(( - 2 h_1 - h_0) h_2 + h_1^2 + h_0 h_1)}{(h_1^2 + h_0 h_1) h_2} \right ) f_{i}^n \nonumber \\
 &- u \tau \frac{u^2 \tau^2 + ( - 2 h_1 - h_0) u \tau + (h_1^2 + h_0 h_1)}{h_2^3 + (2 h_1 + h_0) h_2^2 + (h_1^2 + h_0 h_1) h_2} f_{i + 1}^n
\end{align}
where $h_{0,1}$ are the nearest and next nearest grid spacings in the upwind direction and $h_2$ is the nearest grid spacing in the other direction. 

Note that the Beam--Warming and third order schemes have an upwind bias so for $u < 0$ apply the ``reflection'' transformation: $h_i \rightarrow -h_i$ and $f_{i\pm a} \rightarrow f_{i \mp a}$. 

\bibliographystyle{mn2e}
\bibliography{refs}

\bsp 

\label{lastpage} 

\end{document}

%% file: paperPlots/specIndNonRel.pdf_t
\begin{picture}(0,0)%
\includegraphics{specIndNonRel.pdf}%
\end{picture}%
\setlength{\unitlength}{3947sp}%
\begingroup\makeatletter\ifx\SetFigFontNFSS\undefined%
\gdef\SetFigFontNFSS#1#2#3#4#5{%
  \reset@font\fontsize{#1}{#2pt}%
  \fontfamily{#3}\fontseries{#4}\fontshape{#5}%
  \selectfont}%
\fi\endgroup%
\begin{picture}(8310,6476)(1695,-7366)
%  Begin plot #2 
\put(9078,-1401){\makebox(0,0)[rb]{\smash{{\SetFigFontNFSS{14}{16.8}{\familydefault}{\mddefault}{\updefault}Numerical Results}}}}
\put(2344,-6589){\makebox(0,0)[rb]{\smash{{\SetFigFontNFSS{14}{16.8}{\familydefault}{\mddefault}{\updefault}4}}}}
\put(2344,-5256){\makebox(0,0)[rb]{\smash{{\SetFigFontNFSS{14}{16.8}{\familydefault}{\mddefault}{\updefault}4.5}}}}
\put(2344,-3922){\makebox(0,0)[rb]{\smash{{\SetFigFontNFSS{14}{16.8}{\familydefault}{\mddefault}{\updefault}5}}}}
\put(2344,-2589){\makebox(0,0)[rb]{\smash{{\SetFigFontNFSS{14}{16.8}{\familydefault}{\mddefault}{\updefault}5.5}}}}
\put(2344,-1256){\makebox(0,0)[rb]{\smash{{\SetFigFontNFSS{14}{16.8}{\familydefault}{\mddefault}{\updefault}6}}}}
\put(2449,-7031){\makebox(0,0)[b]{\smash{{\SetFigFontNFSS{14}{16.8}{\familydefault}{\mddefault}{\updefault}2}}}}
\put(4309,-7031){\makebox(0,0)[b]{\smash{{\SetFigFontNFSS{14}{16.8}{\familydefault}{\mddefault}{\updefault}2.5}}}}
\put(6169,-7031){\makebox(0,0)[b]{\smash{{\SetFigFontNFSS{14}{16.8}{\familydefault}{\mddefault}{\updefault}3}}}}
\put(8028,-7031){\makebox(0,0)[b]{\smash{{\SetFigFontNFSS{14}{16.8}{\familydefault}{\mddefault}{\updefault}3.5}}}}
\put(9888,-7031){\makebox(0,0)[b]{\smash{{\SetFigFontNFSS{14}{16.8}{\familydefault}{\mddefault}{\updefault}4}}}}
\put(1879,-3836){\rotatebox{90.0}{\makebox(0,0)[b]{\smash{{\SetFigFontNFSS{17.5}{16.8}{\familydefault}{\mddefault}{\updefault}Spectral Index}}}}}
\put(6168,-7293){\makebox(0,0)[b]{\smash{{\SetFigFontNFSS{17.5}{16.8}{\familydefault}{\mddefault}{\updefault}Compression Ratio}}}}
%  Begin plot #1 
\put(9078,-1151){\makebox(0,0)[rb]{\smash{{\SetFigFontNFSS{14}{16.8}{\familydefault}{\mddefault}{\updefault}$3 \sigma / ( \sigma - 1 )$}}}}
\end{picture}%

%% file: paperPlots/H+M.pdf_t
\begin{picture}(0,0)%
\includegraphics{H+M.pdf}%
\end{picture}%
\setlength{\unitlength}{3947sp}%
\begingroup\makeatletter\ifx\SetFigFontNFSS\undefined%
\gdef\SetFigFontNFSS#1#2#3#4#5{%
  \reset@font\fontsize{#1}{#2pt}%
  \fontfamily{#3}\fontseries{#4}\fontshape{#5}%
  \selectfont}%
\fi\endgroup%
\begin{picture}(8005,6317)(1895,-7207)
%  Begin plot #3 
\put(9288,-1501){\makebox(0,0)[rb]{\smash{{\SetFigFontNFSS{12}{12.0}{\familydefault}{\mddefault}{\updefault}$f_{down} Numerical$}}}}
%  Begin plot #4 
\put(9288,-1701){\makebox(0,0)[rb]{\smash{{\SetFigFontNFSS{12}{12.0}{\familydefault}{\mddefault}{\updefault}$f_{down} Analytical$}}}}
%  Begin plot #2 
\put(9288,-1301){\makebox(0,0)[rb]{\smash{{\SetFigFontNFSS{12}{12.0}{\familydefault}{\mddefault}{\updefault}$f_s Analytical$}}}}
\put(2374,-6564){\makebox(0,0)[rb]{\smash{{\SetFigFontNFSS{14}{12.0}{\familydefault}{\mddefault}{\updefault}0}}}}
\put(2374,-5498){\makebox(0,0)[rb]{\smash{{\SetFigFontNFSS{14}{12.0}{\familydefault}{\mddefault}{\updefault}0.2}}}}
\put(2374,-4431){\makebox(0,0)[rb]{\smash{{\SetFigFontNFSS{14}{12.0}{\familydefault}{\mddefault}{\updefault}0.4}}}}
\put(2374,-3364){\makebox(0,0)[rb]{\smash{{\SetFigFontNFSS{14}{12.0}{\familydefault}{\mddefault}{\updefault}0.6}}}}
\put(2374,-2297){\makebox(0,0)[rb]{\smash{{\SetFigFontNFSS{14}{12.0}{\familydefault}{\mddefault}{\updefault}0.8}}}}
\put(2374,-1231){\makebox(0,0)[rb]{\smash{{\SetFigFontNFSS{14}{12.0}{\familydefault}{\mddefault}{\updefault}1}}}}
\put(3276,-7000){\makebox(0,0)[b]{\smash{{\SetFigFontNFSS{14}{12.0}{\familydefault}{\mddefault}{\updefault}-2}}}}
\put(4653,-7000){\makebox(0,0)[b]{\smash{{\SetFigFontNFSS{14}{12.0}{\familydefault}{\mddefault}{\updefault}-1.5}}}}
\put(6031,-7000){\makebox(0,0)[b]{\smash{{\SetFigFontNFSS{14}{12.0}{\familydefault}{\mddefault}{\updefault}-1}}}}
\put(7408,-7000){\makebox(0,0)[b]{\smash{{\SetFigFontNFSS{14}{12.0}{\familydefault}{\mddefault}{\updefault}-0.5}}}}
\put(8786,-7000){\makebox(0,0)[b]{\smash{{\SetFigFontNFSS{14}{12.0}{\familydefault}{\mddefault}{\updefault}0}}}}
\put(1842,-3836){\rotatebox{90.0}{\makebox(0,0)[b]{\smash{{\SetFigFontNFSS{17.5}{12.0}{\familydefault}{\mddefault}{\updefault}$f$}}}}}
\put(6168,-7243){\makebox(0,0)[b]{\smash{{\SetFigFontNFSS{17.5}{12.0}{\familydefault}{\mddefault}{\updefault}$\log(p/p_0)$}}}}
%  Begin plot #1 
\put(9288,-1101){\makebox(0,0)[rb]{\smash{{\SetFigFontNFSS{12}{12.0}{\familydefault}{\mddefault}{\updefault}$f_s Numerical$}}}}
\end{picture}%

%% file: paperPlots/GUvS.pdf_t
\begin{picture}(0,0)%
\includegraphics{GUvS.pdf}%
\end{picture}%
\setlength{\unitlength}{3947sp}%
\begingroup\makeatletter\ifx\SetFigFontNFSS\undefined%
\gdef\SetFigFontNFSS#1#2#3#4#5{%
  \reset@font\fontsize{#1}{#2pt}%
  \fontfamily{#3}\fontseries{#4}\fontshape{#5}%
  \selectfont}%
\fi\endgroup%
\begin{picture}(5792,3499)(1195,-3999)
%  Begin plot #1 
\put(6375,-711){\makebox(0,0)[rb]{\smash{{\SetFigFontNFSS{10}{12.0}{\familydefault}{\mddefault}{\updefault}Kirk et al}}}}
%  Begin plot #2 
\put(6375,-836){\makebox(0,0)[rb]{\smash{{\SetFigFontNFSS{10}{12.0}{\familydefault}{\mddefault}{\updefault}New Method}}}}
\put(1738,-3388){\makebox(0,0)[rb]{\smash{{\SetFigFontNFSS{12}{12.0}{\familydefault}{\mddefault}{\updefault}4}}}}
\put(1738,-2802){\makebox(0,0)[rb]{\smash{{\SetFigFontNFSS{12}{12.0}{\familydefault}{\mddefault}{\updefault}4.05}}}}
\put(1738,-2216){\makebox(0,0)[rb]{\smash{{\SetFigFontNFSS{12}{12.0}{\familydefault}{\mddefault}{\updefault}4.1}}}}
\put(1738,-1629){\makebox(0,0)[rb]{\smash{{\SetFigFontNFSS{12}{12.0}{\familydefault}{\mddefault}{\updefault}4.15}}}}
\put(1738,-1043){\makebox(0,0)[rb]{\smash{{\SetFigFontNFSS{12}{12.0}{\familydefault}{\mddefault}{\updefault}4.2}}}}
\put(1813,-3748){\makebox(0,0)[b]{\smash{{\SetFigFontNFSS{12}{12.0}{\familydefault}{\mddefault}{\updefault}0}}}}
\put(2752,-3748){\makebox(0,0)[b]{\smash{{\SetFigFontNFSS{12}{12.0}{\familydefault}{\mddefault}{\updefault}2}}}}
\put(3690,-3748){\makebox(0,0)[b]{\smash{{\SetFigFontNFSS{12}{12.0}{\familydefault}{\mddefault}{\updefault}4}}}}
\put(4629,-3748){\makebox(0,0)[b]{\smash{{\SetFigFontNFSS{12}{12.0}{\familydefault}{\mddefault}{\updefault}6}}}}
\put(5567,-3748){\makebox(0,0)[b]{\smash{{\SetFigFontNFSS{12}{12.0}{\familydefault}{\mddefault}{\updefault}8}}}}
\put(6506,-3748){\makebox(0,0)[b]{\smash{{\SetFigFontNFSS{12}{12.0}{\familydefault}{\mddefault}{\updefault}10}}}}
\put(1331,-2037){\rotatebox{90.0}{\makebox(0,0)[b]{\smash{{\SetFigFontNFSS{12}{12.0}{\familydefault}{\mddefault}{\updefault}Spectral Index}}}}}
\put(4394,-3935){\makebox(0,0)[b]{\smash{{\SetFigFontNFSS{12}{12.0}{\familydefault}{\mddefault}{\updefault}$\Gamma_1 u_1$}}}}
\end{picture}%

%% file: paperPlots/WvS.pdf_t
\begin{picture}(0,0)%
\includegraphics{WvS.pdf}%
\end{picture}%
\setlength{\unitlength}{3947sp}%
\begingroup\makeatletter\ifx\SetFigFontNFSS\undefined%
\gdef\SetFigFontNFSS#1#2#3#4#5{%
  \reset@font\fontsize{#1}{#2pt}%
  \fontfamily{#3}\fontseries{#4}\fontshape{#5}%
  \selectfont}%
\fi\endgroup%
\begin{picture}(5933,3496)(1195,-3950)
%  Begin plot #1 
\put(6375,-711){\makebox(0,0)[rb]{\smash{{\SetFigFontNFSS{10}{12.0}{\familydefault}{\mddefault}{\updefault}Schneider \& Kirk '89}}}}
%  Begin plot #2 
\put(6375,-836){\makebox(0,0)[rb]{\smash{{\SetFigFontNFSS{10}{12.0}{\familydefault}{\mddefault}{\updefault}New Method}}}}
\put(1663,-3623){\makebox(0,0)[rb]{\smash{{\SetFigFontNFSS{12}{12.0}{\familydefault}{\mddefault}{\updefault}4.5}}}}
\put(1663,-3242){\makebox(0,0)[rb]{\smash{{\SetFigFontNFSS{12}{12.0}{\familydefault}{\mddefault}{\updefault}5}}}}
\put(1663,-2861){\makebox(0,0)[rb]{\smash{{\SetFigFontNFSS{12}{12.0}{\familydefault}{\mddefault}{\updefault}5.5}}}}
\put(1663,-2480){\makebox(0,0)[rb]{\smash{{\SetFigFontNFSS{12}{12.0}{\familydefault}{\mddefault}{\updefault}6}}}}
\put(1663,-2098){\makebox(0,0)[rb]{\smash{{\SetFigFontNFSS{12}{12.0}{\familydefault}{\mddefault}{\updefault}6.5}}}}
\put(1663,-1717){\makebox(0,0)[rb]{\smash{{\SetFigFontNFSS{12}{12.0}{\familydefault}{\mddefault}{\updefault}7}}}}
\put(1663,-1336){\makebox(0,0)[rb]{\smash{{\SetFigFontNFSS{12}{12.0}{\familydefault}{\mddefault}{\updefault}7.5}}}}
\put(1663,-955){\makebox(0,0)[rb]{\smash{{\SetFigFontNFSS{12}{12.0}{\familydefault}{\mddefault}{\updefault}8}}}}
\put(1663,-574){\makebox(0,0)[rb]{\smash{{\SetFigFontNFSS{12}{12.0}{\familydefault}{\mddefault}{\updefault}8.5}}}}
\put(1738,-3748){\makebox(0,0)[b]{\smash{{\SetFigFontNFSS{12}{12.0}{\familydefault}{\mddefault}{\updefault}0}}}}
\put(2320,-3748){\makebox(0,0)[b]{\smash{{\SetFigFontNFSS{12}{12.0}{\familydefault}{\mddefault}{\updefault}0.05}}}}
\put(2902,-3748){\makebox(0,0)[b]{\smash{{\SetFigFontNFSS{12}{12.0}{\familydefault}{\mddefault}{\updefault}0.1}}}}
\put(3484,-3748){\makebox(0,0)[b]{\smash{{\SetFigFontNFSS{12}{12.0}{\familydefault}{\mddefault}{\updefault}0.15}}}}
\put(4066,-3748){\makebox(0,0)[b]{\smash{{\SetFigFontNFSS{12}{12.0}{\familydefault}{\mddefault}{\updefault}0.2}}}}
\put(4647,-3748){\makebox(0,0)[b]{\smash{{\SetFigFontNFSS{12}{12.0}{\familydefault}{\mddefault}{\updefault}0.25}}}}
\put(5229,-3748){\makebox(0,0)[b]{\smash{{\SetFigFontNFSS{12}{12.0}{\familydefault}{\mddefault}{\updefault}0.3}}}}
\put(5811,-3748){\makebox(0,0)[b]{\smash{{\SetFigFontNFSS{12}{12.0}{\familydefault}{\mddefault}{\updefault}0.35}}}}
\put(6393,-3748){\makebox(0,0)[b]{\smash{{\SetFigFontNFSS{12}{12.0}{\familydefault}{\mddefault}{\updefault}0.4}}}}
\put(6975,-3748){\makebox(0,0)[b]{\smash{{\SetFigFontNFSS{12}{12.0}{\familydefault}{\mddefault}{\updefault}0.45}}}}
\put(1331,-2037){\rotatebox{90.0}{\makebox(0,0)[b]{\smash{{\SetFigFontNFSS{12}{12.0}{\familydefault}{\mddefault}{\updefault}Spectral Index}}}}}
\put(4356,-3935){\makebox(0,0)[b]{\smash{{\SetFigFontNFSS{12}{12.0}{\familydefault}{\mddefault}{\updefault}Shock Width}}}}
\end{picture}%

%% file: paperPlots/FvMu.pdf_t
\begin{picture}(0,0)%
\includegraphics{FvMu.pdf}%
\end{picture}%
\setlength{\unitlength}{3947sp}%
\begingroup\makeatletter\ifx\SetFigFontNFSS\undefined%
\gdef\SetFigFontNFSS#1#2#3#4#5{%
  \reset@font\fontsize{#1}{#2pt}%
  \fontfamily{#3}\fontseries{#4}\fontshape{#5}%
  \selectfont}%
\fi\endgroup%
\begin{picture}(8268,6538)(1696,-7362)
%  Begin plot #1 
\put(9078,-1151){\makebox(0,0)[rb]{\smash{{\SetFigFontNFSS{14}{16.8}{\familydefault}{\mddefault}{\updefault}New Method}}}}
%  Begin plot #2 
\put(9078,-1326){\makebox(0,0)[rb]{\smash{{\SetFigFontNFSS{14}{16.8}{\familydefault}{\mddefault}{\updefault}Kirk et al.}}}}
%  Begin plot #3 
\put(9078,-1501){\makebox(0,0)[rb]{\smash{{\SetFigFontNFSS{14}{16.8}{\familydefault}{\mddefault}{\updefault}Dempsey \& Kirk}}}}
\put(2344,-6856){\makebox(0,0)[rb]{\smash{{\SetFigFontNFSS{14}{16.8}{\familydefault}{\mddefault}{\updefault}0}}}}
\put(2344,-6018){\makebox(0,0)[rb]{\smash{{\SetFigFontNFSS{14}{16.8}{\familydefault}{\mddefault}{\updefault}0.2}}}}
\put(2344,-5180){\makebox(0,0)[rb]{\smash{{\SetFigFontNFSS{14}{16.8}{\familydefault}{\mddefault}{\updefault}0.4}}}}
\put(2344,-4342){\makebox(0,0)[rb]{\smash{{\SetFigFontNFSS{14}{16.8}{\familydefault}{\mddefault}{\updefault}0.6}}}}
\put(2344,-3503){\makebox(0,0)[rb]{\smash{{\SetFigFontNFSS{14}{16.8}{\familydefault}{\mddefault}{\updefault}0.8}}}}
\put(2344,-2665){\makebox(0,0)[rb]{\smash{{\SetFigFontNFSS{14}{16.8}{\familydefault}{\mddefault}{\updefault}1}}}}
\put(2344,-1827){\makebox(0,0)[rb]{\smash{{\SetFigFontNFSS{14}{16.8}{\familydefault}{\mddefault}{\updefault}1.2}}}}
\put(2344,-989){\makebox(0,0)[rb]{\smash{{\SetFigFontNFSS{14}{16.8}{\familydefault}{\mddefault}{\updefault}1.4}}}}
\put(2449,-7031){\makebox(0,0)[b]{\smash{{\SetFigFontNFSS{14}{16.8}{\familydefault}{\mddefault}{\updefault}-1}}}}
\put(4309,-7031){\makebox(0,0)[b]{\smash{{\SetFigFontNFSS{14}{16.8}{\familydefault}{\mddefault}{\updefault}-0.5}}}}
\put(6169,-7031){\makebox(0,0)[b]{\smash{{\SetFigFontNFSS{14}{16.8}{\familydefault}{\mddefault}{\updefault}0}}}}
\put(8028,-7031){\makebox(0,0)[b]{\smash{{\SetFigFontNFSS{14}{16.8}{\familydefault}{\mddefault}{\updefault}0.5}}}}
\put(9888,-7031){\makebox(0,0)[b]{\smash{{\SetFigFontNFSS{14}{16.8}{\familydefault}{\mddefault}{\updefault}1}}}}
\put(1879,-3836){\rotatebox{90.0}{\makebox(0,0)[b]{\smash{{\SetFigFontNFSS{17.5}{16.8}{\familydefault}{\mddefault}{\updefault}$f_s(\mu)$}}}}}
\put(6168,-7293){\makebox(0,0)[b]{\smash{{\SetFigFontNFSS{17.5}{16.8}{\familydefault}{\mddefault}{\updefault}$\mu$}}}}
\end{picture}%

%% file: paperPlots/FvMu2.pdf_t
\begin{picture}(0,0)%
\includegraphics{FvMu2.pdf}%
\end{picture}%
\setlength{\unitlength}{3947sp}%
\begingroup\makeatletter\ifx\SetFigFontNFSS\undefined%
\gdef\SetFigFontNFSS#1#2#3#4#5{%
  \reset@font\fontsize{#1}{#2pt}%
  \fontfamily{#3}\fontseries{#4}\fontshape{#5}%
  \selectfont}%
\fi\endgroup%
\begin{picture}(8268,6538)(1696,-7362)
%  Begin plot #1 
\put(9078,-1151){\makebox(0,0)[rb]{\smash{{\SetFigFontNFSS{14}{16.8}{\familydefault}{\mddefault}{\updefault}New Method}}}}
%  Begin plot #2 
\put(9078,-1326){\makebox(0,0)[rb]{\smash{{\SetFigFontNFSS{14}{16.8}{\familydefault}{\mddefault}{\updefault}Kirk et al}}}}
%  Begin plot #3 
\put(9078,-1501){\makebox(0,0)[rb]{\smash{{\SetFigFontNFSS{14}{16.8}{\familydefault}{\mddefault}{\updefault}Dempsey and Kirk}}}}
\put(2344,-6856){\makebox(0,0)[rb]{\smash{{\SetFigFontNFSS{14}{16.8}{\familydefault}{\mddefault}{\updefault}0}}}}
\put(2344,-5389){\makebox(0,0)[rb]{\smash{{\SetFigFontNFSS{14}{16.8}{\familydefault}{\mddefault}{\updefault}0.5}}}}
\put(2344,-3922){\makebox(0,0)[rb]{\smash{{\SetFigFontNFSS{14}{16.8}{\familydefault}{\mddefault}{\updefault}1}}}}
\put(2344,-2456){\makebox(0,0)[rb]{\smash{{\SetFigFontNFSS{14}{16.8}{\familydefault}{\mddefault}{\updefault}1.5}}}}
\put(2344,-989){\makebox(0,0)[rb]{\smash{{\SetFigFontNFSS{14}{16.8}{\familydefault}{\mddefault}{\updefault}2}}}}
\put(2449,-7031){\makebox(0,0)[b]{\smash{{\SetFigFontNFSS{14}{16.8}{\familydefault}{\mddefault}{\updefault}-1}}}}
\put(4309,-7031){\makebox(0,0)[b]{\smash{{\SetFigFontNFSS{14}{16.8}{\familydefault}{\mddefault}{\updefault}-0.5}}}}
\put(6169,-7031){\makebox(0,0)[b]{\smash{{\SetFigFontNFSS{14}{16.8}{\familydefault}{\mddefault}{\updefault}0}}}}
\put(8028,-7031){\makebox(0,0)[b]{\smash{{\SetFigFontNFSS{14}{16.8}{\familydefault}{\mddefault}{\updefault}0.5}}}}
\put(9888,-7031){\makebox(0,0)[b]{\smash{{\SetFigFontNFSS{14}{16.8}{\familydefault}{\mddefault}{\updefault}1}}}}
\put(1879,-3836){\rotatebox{90.0}{\makebox(0,0)[b]{\smash{{\SetFigFontNFSS{17.5}{16.8}{\familydefault}{\mddefault}{\updefault}$f_s(\mu)$}}}}}
\put(6168,-7293){\makebox(0,0)[b]{\smash{{\SetFigFontNFSS{17.5}{16.8}{\familydefault}{\mddefault}{\updefault}$\mu$}}}}
\end{picture}%

%% file: paperPlots/relCut.pdf_t
\begin{picture}(0,0)%
\includegraphics{relCut.pdf}%
\end{picture}%
\setlength{\unitlength}{3947sp}%
\begingroup\makeatletter\ifx\SetFigFontNFSS\undefined%
\gdef\SetFigFontNFSS#1#2#3#4#5{%
  \reset@font\fontsize{#1}{#2pt}%
  \fontfamily{#3}\fontseries{#4}\fontshape{#5}%
  \selectfont}%
\fi\endgroup%
\begin{picture}(8268,6816)(1696,-7371)
\put(2344,-6856){\makebox(0,0)[rb]{\smash{{\SetFigFontNFSS{14}{16.8}{\familydefault}{\mddefault}{\updefault}0}}}}
\put(2344,-5789){\makebox(0,0)[rb]{\smash{{\SetFigFontNFSS{14}{16.8}{\familydefault}{\mddefault}{\updefault}0.2}}}}
\put(2344,-4723){\makebox(0,0)[rb]{\smash{{\SetFigFontNFSS{14}{16.8}{\familydefault}{\mddefault}{\updefault}0.4}}}}
\put(2344,-3656){\makebox(0,0)[rb]{\smash{{\SetFigFontNFSS{14}{16.8}{\familydefault}{\mddefault}{\updefault}0.6}}}}
\put(2344,-2589){\makebox(0,0)[rb]{\smash{{\SetFigFontNFSS{14}{16.8}{\familydefault}{\mddefault}{\updefault}0.8}}}}
\put(2344,-1522){\makebox(0,0)[rb]{\smash{{\SetFigFontNFSS{14}{16.8}{\familydefault}{\mddefault}{\updefault}1}}}}
\put(2449,-7031){\makebox(0,0)[b]{\smash{{\SetFigFontNFSS{14}{16.8}{\familydefault}{\mddefault}{\updefault}-5}}}}
\put(3937,-7031){\makebox(0,0)[b]{\smash{{\SetFigFontNFSS{14}{16.8}{\familydefault}{\mddefault}{\updefault}-4}}}}
\put(5425,-7031){\makebox(0,0)[b]{\smash{{\SetFigFontNFSS{14}{16.8}{\familydefault}{\mddefault}{\updefault}-3}}}}
\put(6912,-7031){\makebox(0,0)[b]{\smash{{\SetFigFontNFSS{14}{16.8}{\familydefault}{\mddefault}{\updefault}-2}}}}
\put(8400,-7031){\makebox(0,0)[b]{\smash{{\SetFigFontNFSS{14}{16.8}{\familydefault}{\mddefault}{\updefault}-1}}}}
\put(9888,-7031){\makebox(0,0)[b]{\smash{{\SetFigFontNFSS{14}{16.8}{\familydefault}{\mddefault}{\updefault}0}}}}
\put(1879,-3836){\rotatebox{90.0}{\makebox(0,0)[b]{\smash{{\SetFigFontNFSS{17.5}{16.8}{\familydefault}{\mddefault}{\updefault}$p^s f$}}}}}
\put(6168,-7293){\makebox(0,0)[b]{\smash{{\SetFigFontNFSS{17.5}{16.8}{\familydefault}{\mddefault}{\updefault}$\log(p/p_0)$}}}}
\put(6168,-726){\makebox(0,0)[b]{\smash{{\SetFigFontNFSS{17.5}{16.8}{\familydefault}{\mddefault}{\updefault}$u_1 = 0.995c$, $D = const$, $s = 4.28$}}}}
%  Begin plot #1 
\put(9078,-1151){\makebox(0,0)[rb]{\smash{{\SetFigFontNFSS{14}{16.8}{\familydefault}{\mddefault}{\updefault}New Method}}}}
%  Begin plot #3 
\put(9076,-1711){\makebox(0,0)[rb]{\smash{{\SetFigFontNFSS{14}{16.8}{\familydefault}{\mddefault}{\updefault}$\exp( -(p/p_{fit})^2 )$}}}}
%  Begin plot #2 
\put(9076,-1411){\makebox(0,0)[rb]{\smash{{\SetFigFontNFSS{14}{16.8}{\familydefault}{\mddefault}{\updefault}$\exp( -(p/p_{fit}) )$}}}}
\end{picture}%